\documentclass[amsmath,amssymb,twocolumn]{revtex4}
\usepackage{epsfig}
\usepackage{graphicx}
\usepackage{color}
\usepackage{verbatim}

\begin{document}

\title{Spin-Boson Model with Diagonal and Off-Diagonal Coupling to Two Independent Baths: Ground-State Phase Transition in the Deep Sub-Ohmic Regime}

\author{Yang Zhao$^{1}$, Yao Yao$^{2}$, Vladimir Chernyak$^{3}$, Yang Zhao$^{1}$\footnote{Electronic address:~\url{YZhao@ntu.edu.sg}}}
\date{\today}
\address{$^1$Division of Materials Science, Nanyang Technological University, Singapore 639798, Singapore\\
$^2$State Key Laboratory of Surface Physics and Department of Physics, Fudan University,
Shanghai 200433, China\\
$^3$Department of Chemistry, Wayne State University, Detroit, Michigan 48202, United States
}

\begin{abstract}
     We investigate a spin-boson model with two boson baths that are coupled to two perpendicular components of the spin by employing the density matrix renormalization group method with an optimized boson basis. It is revealed that in the deep sub-Ohmic regime there exists a novel second-order phase transition between two types of doubly degenerate states, which is reduced to one of the usual type for nonzero tunneling. In addition, it is found that expectation values of the spin components display jumps at the phase boundary in the absence of bias and tunneling.
\end{abstract}

\maketitle

The spin boson model (SBM) \cite{Leggett,Weiss} describes a two-level system coupled to a bosonic bath often represented by a set of harmonic oscillators. As an archetype model for quantum dissipation, the SBM has been widely used in fields such as quantum computation \cite{qs1,qs2} and qubit decoherence \cite{dc1}, amorphous solids \cite{amo}, biological molecules \cite{et1,et2}, as well as studies of thermodynamic properties \cite{ther}, spin dynamics \cite{Leggett,dyna} and quantum phase transitions \cite{qpt1,qpt2}. The SBM can be written as
\begin{eqnarray}\label{sbm1_h}
 \textrm{H}_{\textrm{SBM}}&=&\frac{\varepsilon}{2}\sigma_z-\frac{\Delta}{2}\sigma_x+\sum_l \omega_l b_l^\dag b_l\nonumber\\
\label{Ohami}
&+&\frac{\sigma_z}{2}\sum_l \lambda_l(b^\dag_l+b_l),
\end{eqnarray}
where $\varepsilon$ is the spin bias, $\sigma_{x}$ and $\sigma_z$ are pauli matrices, $\Delta$ is the tunneling constant, $\omega_l$ is the frequency of the $l$-th boson mode for which $b_l$($b^{\dagger}_{l}$) denotes the boson annihilation (creation) operator, and $\lambda_l$ signifies the coupling amplitude with the spin.
For a quasi-continuous spectral density function $J(\omega)\equiv \sum_l \lambda^2_l \delta(\omega-\omega_l)$,  a power law form can be adopted in the low-frequency regime: $J(\omega) = 2\pi \alpha\omega_{c}^{1-s}\omega^{s}$, 
where $\omega_c$ is the cut-off frequency, $\alpha$ is the spin-bath coupling constant, and $s$ is the spectral exponent characterizing bath properties so that $s=1$ and $s<1$ ($s>1$) are known as the Ohmic and sub-Ohmic (super-Ohmic) regime, respectively.
Studies \cite{qpt1,qpt2} have shown that if $\varepsilon=0$ and $s<1$, strong spin-bath coupling induces spontaneous symmetry breaking restricting the orientation of the spin-$1/2$ to a specific direction (spin-up or down). Thus, the spin-$1/2$ will be in a two-fold degenerate state, and the entire system, described by Eq.~(\ref{sbm1_h}), is said to be in the 'localized' phase. For weak coupling, the spin is free to flip between the spin-up and the spin-down states, and the system is in the 'delocalized' phase. A critical coupling strength $\alpha_c$ exists for this second order phase transition, which for $s=1$ emerges as a Kosterlitz-Thouless transition \cite{ohmic}.

One can add to Hamiltonian (\ref{sbm1_h}) an off-diagonal coupling term, represented by $\sigma_z/2\sum_l \bar{\lambda}_l(b^\dag_l+b_l)$. Recent studies \cite{lv} reveal that in the sub-Ohmic regime, the off-diagonal interaction could lift the degeneracy in the localized phase, hence removing the second order phase transition, while there may exist a first order phase transition when the diagonal and off-diagonal coupling strengths are chosen properly.
To obtain deeper understandings on the competition between the diagonal and off-diagonal coupling, an additional boson bath, coupled to the spin-$1/2$ off-diagonally, is taken into consideration, resulting in a so-called ``two-bath SBM."
In the limit of zero tunneling, the model possesses a high level of symmetry corresponding to a non-abelian group that contains eight elements. Our symmetry-based analysis shows that the system ground state is always doubly degenerate, and the phase transition occurs not between phases with degenerate and non-degenerate ground states, but rather due to the fact that ground-state degeneracy does not necessarily imply spontaneous symmetry breaking. Stated differently, a special type of quantum phase transitions is identified here, which is confirmed by results from the density matrix renormalization group (DMRG) calculations, a method that has been proven in numerous studies of quantum phase transitions in the usual SBM \cite{guo}.

Previous studies, such as DMRG, numerical renormalization group (NRG), quantum Monte Carlo (QMC) and variational methods, have revealed that in the absence of bias $\langle\sigma_z\rangle$ will be zero if $\alpha$ is below some critical value $\alpha_c(\Delta)$, implying the system to be in a delocalized phase. If $\alpha >\alpha_c$, $\langle\sigma_z\rangle$ acquires a finite value and the system enters into a localized phase. This well known delocalized-localized transition is ascribed to the competition between the spin-bath coupling and the tunneling constant.
Off-diagonal coupling between spin and the boson bath, which can be described by  $\frac{\sigma_x}{2}\sqrt{\frac{\eta}{\pi}}(b^\dagger_0+b_0)$, provides an alternative channel of communications between spin down $|\downarrow\rangle$ and up $|\uparrow\rangle$ states. The single-bath SBM has been investigated via the Davydov D$_1$ variational ansatz \cite{lv}, and a novel first order phase transition was found to arise when the off-diagonal coupling is taken into account along with the diagonal coupling.
Motivated by this finding, we expect much richer ground state properties can be uncovered when the diagonal and the off-diagonal coupling is ascribed to two boson baths rather than a common one. The Hamiltonian for the two-bath SBM can be given as
\begin{eqnarray}
\hat{H}&=&\frac{\varepsilon}{2}\sigma_z-\frac{\Delta}{2}\sigma_x+\sum_{l,i} \omega_l b_{l,i}^\dag b_{l,i}\nonumber\\
\label{Ohami}
&+&\frac{\sigma_z}{2}\sum_l \lambda_l(b^\dag_{l,1}+b_{l,1})+\frac{\sigma_x}{2}\sum_l \phi_l(b^\dag_{l,2}+b_{l,2}),
\end{eqnarray}
where the subscript $i=1,2$ is introduced to distinguish the two baths, and $\lambda_l$ and $\phi_l$ are the diagonal and off-diagonal coupling strengths, respectively, which can be used to determine spectral densities,
\begin{eqnarray}\label{OspectraZ}
J_z(\omega)=\sum_l \lambda^2_l \delta(\omega-\omega_l)\Rightarrow2\alpha\omega_c^{1-s}\omega^s,\\
\label{OspectraX}
J_x(\omega)=\sum_l \phi^2_l \delta(\omega-\omega_l)\Rightarrow2\beta\omega_c^{1-\bar{s}}\omega^{\bar{s}}.
\end{eqnarray}
Here, $\alpha$ and $\beta$ are dimensionless coupling constants, and $\omega_c$ is set to be unity throughout this work. The two baths are characterized by the spectral exponents $s$ and $\bar{s}$. 

Eq.~(\ref{sbm1_h}) can be recast into its continuum form
\begin{eqnarray}\label{sbm1_ctnu_h}
 \textrm{H}_{\textrm{SBM}}&=&\frac{\varepsilon}{2}\sigma_z-\frac{\Delta}{2}\sigma_x+\int^{\omega_c}_{0} g(\omega)b_{\omega}^\dag b_{\omega}\nonumber\\
\label{Ohami}
&+&\frac{\sigma_z}{2}\int^{\omega_c}_{0}h(\omega)(b^\dag_{\omega}+b_{\omega}),
\end{eqnarray}
where $b_{\omega}$ and $b^\dagger_{\omega}$ are the counterparts of $b_{l}$ and $b^\dagger_{l}$, $g(\omega)$ is the dispersion relation, and $h(\omega)$ is the coupling function. As indicated in Refs.~\cite{chinmap} and \cite{qpt2}, $g(\omega)$ and $h(\omega)$ obey
\begin{equation}\label{gh}
  \textrm{J}(\omega) = \pi\frac{dg^{-1}(\omega)}{d\omega}h^2(g^{-1}(\omega)),
\end{equation}
with $g^{-1}(\omega)$ being the inverse function of $g(\omega)$. Starting  from Eq.~(\ref{sbm1_ctnu_h}), and using the canonical transformation\cite{Wilson,chinmap}, we can map the boson bath onto a Wilson chain, with Eq.~(\ref{sbm1_ctnu_h}) being mapped simultaneously onto
\begin{eqnarray}
\hat{H}&=&\frac{\varepsilon}{2}\sigma_z-\frac{\Delta}{2}\sigma_x+\frac{\sigma_z}{2}\sqrt{\frac{\eta}{\pi}}(p^\dag_0+p_0)\nonumber\\
&+&\sum_{n=0} [\omega_n p_n^\dag p_n + t_n(p_n^\dag p_{n+1}+p_{n+1}^\dag p_n)],\label{wil_sbm1}
\end{eqnarray}
here $p_n^\dag$ ($p_n$) are boson creation (annihilation) operator, $\omega_n$ is the on site energy of site $n$, $t_n$ is the hopping amplitude, the coupling constant $\eta$ is proportional to $\alpha$.
In order to deal with the two-bath SBM by employing the DMRG algorithm, followed by the standard treatment \cite{Wilson,qpt2,chinmap} that leads to
Eq.~(\ref{wil_sbm1}), the two boson baths are transformed into two Wilson chains. The Hamiltonian (\ref{Ohami}) is mapped simultaneously to\cite{chinmap}:
\begin{eqnarray}
\hat{H}&=&\frac{\varepsilon}{2}\sigma_z-\frac{\Delta}{2}\sigma_x\nonumber\\
&+&\sum_{n=0,i} [\omega_{n,i} p_{n,i}^\dag p_{n,i} + t_{n,i}(p_{n,i}^\dag p_{n+1,i}+p_{n+1,i}^\dag p_{n,i})]\nonumber\\
\label{W_ham}
&+&\frac{\sigma_z}{2}\sqrt{\frac{\eta_z}{\pi}}(p^\dag_{0,1}+p_{0,1})+\frac{\sigma_x}{2}\sqrt{\frac{\eta_x}{\pi}}(p^\dag_{0,2}+p_{0,2}),
\end{eqnarray}
where $i = 1, 2$ label the baths, and
\begin{eqnarray}\label{etax}
  \eta_{x}& = &\int_{0}^{\omega_c}J_{x}(\omega)d\omega=\frac{2\pi\beta}{1+\bar{s}}\omega^2_c,\\
  \eta_{z}& = &\int_{0}^{\omega_c}J_{z}(\omega)d\omega=\frac{2\pi\alpha}{1+s}\omega^2_c, \label{etaz}\\
  \omega_{n,1} &= &\zeta_s(A_n+C_n),~~t_{n,1} = \zeta_s(\frac{N_{n+1}}{N_n})A_n,\\
  \zeta_s &= &\frac{s+1}{s+2}\frac{1-\lambda^{-(s+2)}}{1-\lambda^{-(s+1)}}\omega_c,\nonumber\\
  A_n &=& \lambda^{-n}\frac{(1-\lambda^{-(n+1+s)})^2}{(1-\lambda^{-(2n+1+s)})(1-\lambda^{-(2n+2+s)})},\nonumber\\
  C_n &=& \lambda^{-n+s}\frac{(1-\lambda^{-n})^2}{(1-\lambda^{-(2n+s)})(1-\lambda^{-(2n+1+s)})},\nonumber\\
  N^2_n &= &\lambda^{-n(1+s)}\frac{(\lambda^{-1};\lambda^{-1})_{n}^2}{(\lambda^{-(s+1)};\lambda^{-1})_{n}^2(1-\lambda^{-(2n+1+s)})},\nonumber
\end{eqnarray}
with
$ (a;b)_n = (1-a)(1-ab)(1-ab^2)\cdots(1-ab^{(n-1)})$.
Here $\lambda>1$ is the discretization parameter.
In the Fock representation, the ground state wave function of Hamiltonian (\ref{W_ham}) characterizing a single chain system can be written in the form of matrix-product states (MPS) as
\begin{equation}\label{mps}
  |\psi\rangle = \sum_{i_0 = \uparrow,\downarrow; {j}}X^{i_0}X^{j_1}X^{j_2}\cdots X^{j_{L-1}}|i_0,\vec{j}\rangle,
\end{equation}
where $i_0$ is the spin index, $\vec{j} = (j_1,j_2,\cdots j_{L-1})$, with $0\leq j_i\leq d_p$, represents the quantum numbers for the boson basis, $L$ is the length of the chain (chosen as $51$), and $d_p$ is the the number of boson modes allocated on each site. ${X^j}$ are single matrices whose dimensions are restricted by a cut off $D_c = 50$.
Subsequently, performing an iterative optimization procedure\cite{mps_c}, each matrix $X$ can be optimized to a truncation error less than $10^{-7}$. Furthermore, if a DMRG algorithm with an optimized boson basis\cite{guo} is used, the boson number $d_p$ on each site of the Wilson chain can be kept up to 100. Therefore, a total of $10^2L$ phonons are included in the calculations. A minimum of $d_p=20$ phonons need to be kept
to arrive at a clear conclusion about the phase transition. Using obtained MPS wave functions, we can extract $\langle\sigma_{x}\rangle$, $\langle\sigma_{z}\rangle$ and the von-Neumann entropy $S_{v-N}\equiv -\textrm{Tr}\rho_s\textrm{log}\rho_s$,
where $\rho_s$ is the reduced density matrix of the spin.

The sub-Ohmic SBM with $\beta = 0$ and the spectral density (\ref{OspectraZ}) may exhibit a second order  transition from a delocalized phase ($\langle\sigma_z\rangle = 0 $) to a localized one ($\langle\sigma_z\rangle \neq 0 $), if $\alpha>\alpha_c$ ($0<\alpha_c<1$) \cite{qpt2}. Especially, if $s<1/2$, critical exponents of the phase transition, such as $\langle\sigma_z\rangle = (\alpha-\alpha_c)^{\beta_{\rm MF}} $ where $\beta_{\rm MF}=1/2$, can be obtained via quantum-to-classical correspondence as demonstrated by a variety of numerical techniques \cite{qpt1,qpt2}.
In the two-bath SBM of Eq.~(\ref{W_ham}), competition between the baths pose a significant challenge to the numerical simulations due to an increased total boson number that must be kept. DMRG calculations \cite{guo} have so far revealed that if $s=\bar{s}<1/2$ and $\sigma_{x}$ and $\sigma_{z}$ coupled to two boson baths with equivalent coupling strengths ($\alpha=\beta$), the spin is situated in a localized state. Further, to obtain a deeper insight into the properties of the two-bath SBM, it is interesting to investigate the deep sub-Ohmic regime of the two-bath SBM with different $\alpha$ and $\beta$, for a general scenario of $s=\bar{s}$ and $s\neq\bar{s}$. At last, we will discuss the situations with finite $\varepsilon$ or $\Delta$.

\begin{figure}[tbp]
  \centering
  \vspace{0.3cm}
  \includegraphics[height=!, width=4.5cm, bb = 0 0 277 510]{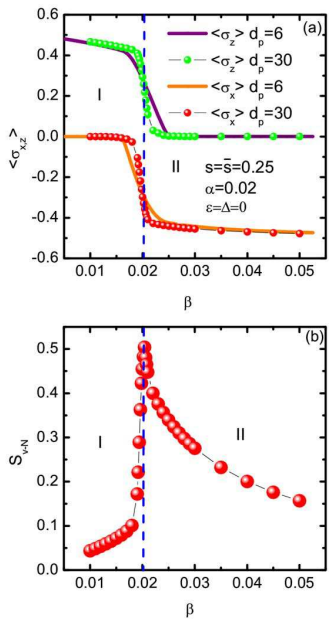}\\
  \caption{ (a) $\langle\sigma_{x}\rangle$ and $\langle\sigma_{z}\rangle$ calculated as a function of $\beta$ using two on-site boson number $d_p = 6$ and $30$; (b) the von-Neumann entropy $S_{v-N}$ as a function of $\beta$. The position of the critical point is labeled by the vertical dashed line, and we set $s=\bar{s}=0.25$ and $\alpha=0.02$. }\label{sbm1}
\end{figure}

We first explore the case of $\varepsilon = \Delta = 0$ and $s=\bar{s}=0.25$ for which Hamiltonian (\ref{W_ham}) is invariant under operation
\begin{equation}\label{parity}
  \mathcal{P} = \sigma_y\textrm{e}^{i\sum_{n}(b^{\dagger}_{n,1}b_{n,1}+b^{\dagger}_{n,2}b_{n,2})},
\end{equation}
indicating a two-fold degeneracy of the ground state. A tiny symmetry-breaking perturbation, a tiny symmetry-breaking perturbation, is often applied to a state with two-fold degeneracy in the DMRG calculations.
Due to diagonal coupling, the spin will be trapped with a finite $\langle\sigma_{z}\rangle$, forming a localized phase. The coupling with $\sigma_x$, however, induces a spin flip between $|\uparrow\rangle$ and $|\downarrow\rangle$, thereby hindering the self-trapping process.
Fig.~\ref{sbm1} (a) shows calculated $\langle\sigma_x\rangle$ and $\langle\sigma_z\rangle$ for $\alpha = 0.02$ and a range of $\beta$ values from 0.0 to 0.05. It is clear that when the off-diagonal coupling is dominant, i.e., $\beta \gg \alpha$, $\langle \sigma_x\rangle$ is finite so that the spin is in the superposition state of $|\uparrow\rangle$ and $|\downarrow\rangle$. We ascribe this phase as 'phase I'. Similar arguments remain valid for the case of $\beta \ll \alpha$, when $\langle \sigma_z\rangle$ assumes a finite value and we term this phase as 'localized phase II', abbreviated as 'phase II'. Further, as shown in Fig.~\ref{sbm1} (b), $S_{v-N}$ also shows a sharp peak at the critical point, $\beta\sim0.0204$. In addition, we have also calculated the fidelity near the critical point reaching the same conclusion. As shown in Fig. \ref{sbm1} (a), $\langle\sigma_z\rangle$ and $\langle\sigma_x\rangle$ are sensitive to the boson number $d_p$ being kept in DMRG calculation. Evidently, to obtain reliable data at the critical point, it is necessary to choose a sufficiently large $d_p$ (over 20).

\begin{figure}[tbp]
  \centering
  \includegraphics[width=0.62   \linewidth]{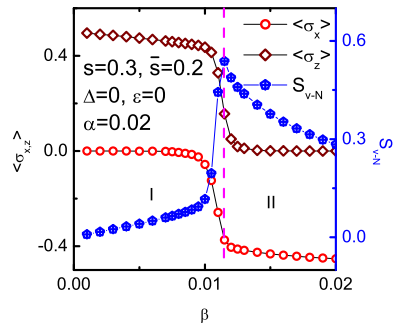}\\
  \caption{ $\langle\sigma_{x}\rangle$ and $\langle\sigma_{z}\rangle$ with respect to $\beta$. Here, $s=0.3$, $\bar{s}=0.2$, and $\alpha=0.02$. The transition points are marked by a pink dashed line. The corresponding von-Neumann entropy ($S_{v-N}$) with respect to $\beta$ is shown as well. Remarkably, $S_{v-N}$ shows a sharp peak at the critical point, at about 0.0115 where phase I goes into II.}\label{sns1}
\end{figure}

Next, we study the case of $s\neq\bar{s}$. According to Eq.~(\ref{etax}) [Eq.~(\ref{etaz})], if $\omega<\omega_c = 1$, the strength of $\eta_x$ ($\eta_y$) is inversely proportional to $1+s$ ($1+\bar{s}$). Therefore, as opposed to the case of $s=\bar{s}$, where the spin-bath interactions are governed solely by $\alpha$ and $\beta$, if $s\neq\bar{s}$, the effective spin-bath interactions are modified, leading to a shift of the two critical points as shown in Fig. \ref{sbm1}.
In Fig.~\ref{sns1}, we present calculated $\langle\sigma_z\rangle$, and $\langle\sigma_x\rangle$ for the case of $s=0.3$, $\bar{s}=0.2$. Similarly, the properties of the transition, from I to II, are analogous to those exhibited in Fig.~\ref{sbm1} (a), the critical point moves from 0.0204 to 0.0115, as indicated by the peak of the entanglement entropy in Fig.~\ref{sns1}. It is convenient to renormalize $\alpha$ and $\beta$ by the factors $1/(1+s)$ and $1/(1+\bar{s})$, respectively. Here, $s$ ($\bar{s}$) increases (decreases) from 0.25 to 0.3 (0.2), therefore, the effective diagonal (off diagonal) coupling will become smaller (larger). In order to reproduce the pha
se transition in Fig.~\ref{sbm1}, the critical value of $beta$ will
have to the left, which is just the result shown in Fig.~\ref{sns1}.

It is now clear that due to the competition of the two baths a second order phase transition exists in the two-bath SBM. In the absence of $\varepsilon$ and $\Delta$, $\sigma_x$ and $\sigma_z$ swap their roles through a rotation along the $y$ axis. This results in a similar swap of $\langle\sigma_x\rangle$ and $\langle\sigma_z\rangle$ near the critical point, where $\langle\sigma_x\rangle$ displays a kink when $\Delta\neq0$. In contrast to the single-bath SBM, $\langle\sigma_x\rangle$ vanishes due to the full $SU$(2) symmetry of the spin and the absence of a confining potential for $\sigma_x$. It should be stressed that both phases, Phases I and II, are doubly degenerate, in agreement with the parity symmetry of Hamiltonian (\ref{W_ham}). The degeneracy of phase I (II) is characterized by the eigenstates of $\sigma_z$ ($\sigma_x$), $|\uparrow\rangle$ and $|\downarrow\rangle$ ($|\leftarrow\rangle$ and $|\rightarrow\rangle$). This is a novel feature of a second order phase transition between states with two-fold degeneracy as a result of bath  competition.

\begin{figure}[tbp]
  \centering
  \includegraphics[width=0.7\linewidth]{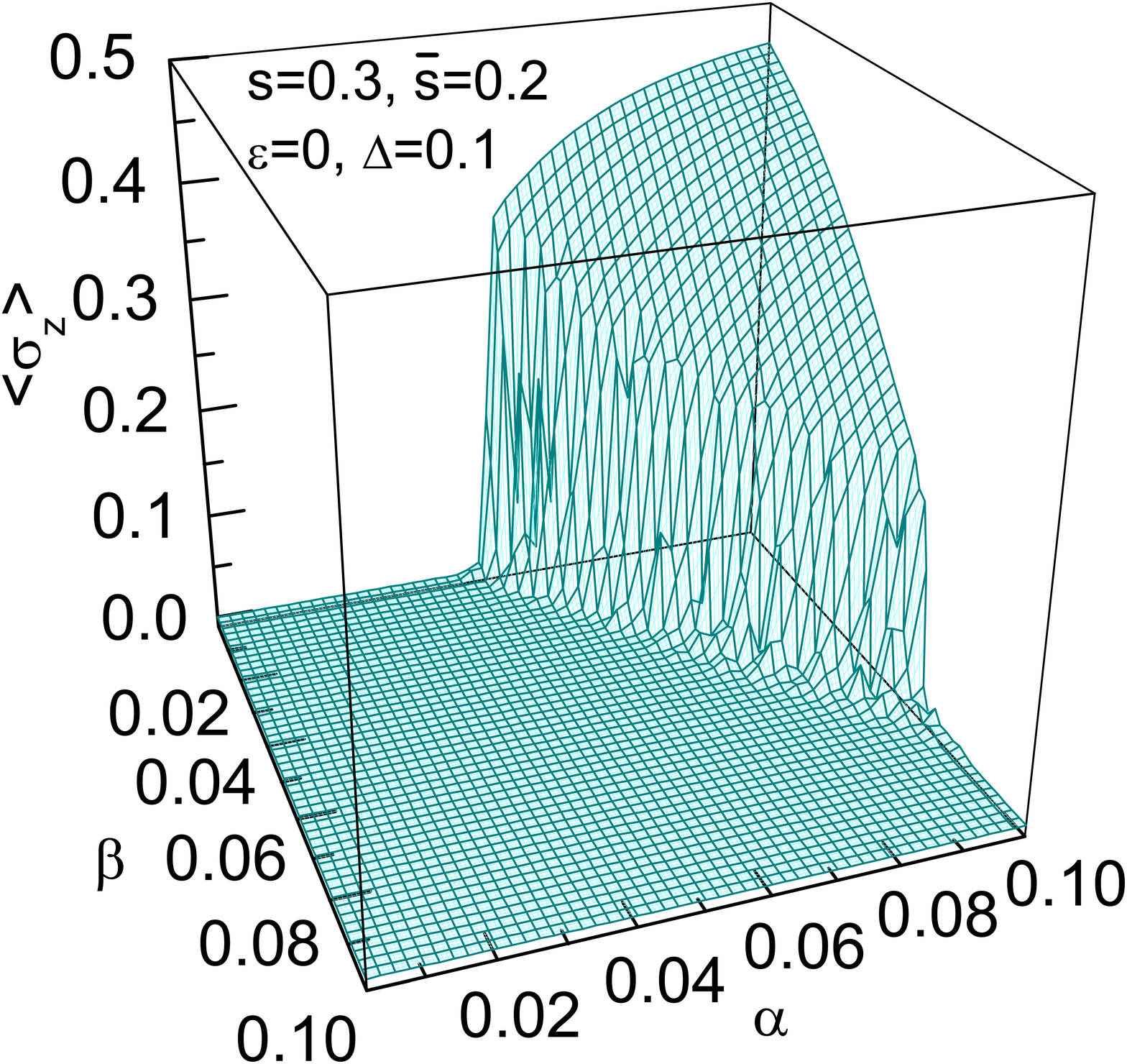}\\
  \caption{$\langle \sigma_{z}\rangle$ as a function of $\alpha$ and $\beta$, wherein $s=0.3$, $\bar{s}=0.2$, $\Delta=0.1$.}\label{phdsz}
\end{figure}

\begin{figure}[tb]
  \centering
  \includegraphics[width=0.62\linewidth]{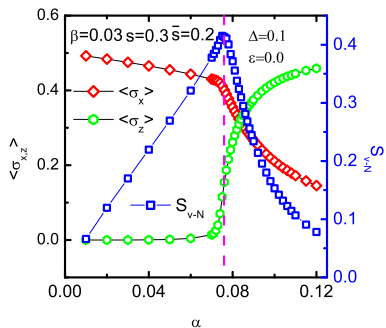}\\
  \caption {$\langle\sigma_z\rangle$, $\langle\sigma_x\rangle$, and entanglement entropy $S_{v-N}$ as a function of $\alpha$ near the critical point, pointing to a second order phase transition. $\langle\sigma_x\rangle$ shows a small kink at the critical point.}\label{Sz}
\end{figure}
 
 As pointed out in Ref.~\cite{lv}, finite $\varepsilon$ or $\Delta$ can break the symmetry of the ground-state free energy and thus prevent the occurrence of a second order phase transition. For $s=0.3$, $\bar{s}=0.2$, Fig.~\ref{phdsz} shows $\langle\sigma_z\rangle$ as a function of $\alpha$ and $\beta$, where $\Delta$ is imposed on the $x$ spin component. Fig.~\ref{Sz}, which displays $\langle\sigma_z\rangle$, $\langle\sigma_x\rangle$, and  $S_{v-N}$ as a function of $\alpha$ for the case of $\beta=0.03$, further confirms the phase boundary in Fig.~\ref{phdsz}. Unlike the large spike at the critical point shown in Fig.~\ref{sbm1}, only a much less pronounced kink is found in $\langle\sigma_x\rangle$. It is argued that the occurrence of the kink is ascribed to a sufficiently large $\Delta$. Similar results can be obtained under a spin bias in the $z$ component after rotating along the $y$ axis. Moreover, through intensive DMRG calculations, we find that $\langle\sigma_z\rangle$ can be reduced to zero by increasing $\Delta$ in the localized phase, while $\langle\sigma_x\rangle$ reaches a saturation value.

To summarize, in the deep sub-Ohmic regime, for an extended SBM with two baths coupled to the $x$ and $z$ spin components, there exists a second order phase transition, from the doubly degenerated 'coherent phase I' to the other doubly degenerated 'localized phase II'. This phase transition, which survives the introduction of finite $\Delta$ or $\varepsilon$, offers a notable difference between the single-bath SBM and the two-bath SBM. Varying bath spectral densities ($s\neq\bar{s}$) shifts the critical point, and for $\varepsilon = 0$ and $\Delta=0$, $\langle\sigma_z\rangle$ and $\langle\sigma_x\rangle$ display jumps near the critical point, a feature that is absent from the single-bath SBM.

\textit{Acknowledgements} - This work is supported by the Singapore National Research Foundation under Project No.~NRF-CRP5-2009-04 and the U.S.~National
Science Foundation under Grant CHE-1111350.

\end{document}